\documentclass{ws-ijmpa}

\begin{document}

\markboth{Satoshi Iso}
{Hawking Radiation and Gravitational Anomaly }

%
\catchline{}{}{}{}{}
%

\title{Hawking Radiation, Gravitational  Anomaly, 
and Conformal Symmetry
-- the Origin of Universality --  
\footnote{Invited talk presented  at the international workshop
"Progress of String Theory and Quantum Field Theory"
at Osaka City University (December 7-10, 2007)}
}

\author{SATOSHI ISO}

\address{
High Energy Accelerator Research Organization (KEK),  \\
and \\
Department of Particles and Nuclear Physics, \\
The Graduate University for Advanced Studies, \\
 Tsukuba, Ibaraki 305-0801, Japan
\\
satoshi.iso@kek.jp}

\maketitle


\begin{abstract}
The universal behavior of Hawking radiation 
is originated in the conformal symmetries of 
matter fields near the black hole horizon.
We explain the origin of this universality
based on (1) the gravitational anomaly 
and its higher-spin generalizations and (2)  
conformal transformation properties of 
fluxes.

\end{abstract}

\ccode{PACS numbers: 04.62.+v, 04.70.Dy, 11.25.Hf, 11.30.-j}

\section{Introduction}	
Hawking radiation is the most prominent quantum effect 
to arise for quantum fields in a background space-time with an event horizon, and if we neglect the grey body factor 
induced by the effect of the scattering outside the horizon,
the flux of the radiation is universally determined 
by the properties of black holes at the horizon.

We will explain that this universality
can be understood   in terms of the gravitational
anomalies at the horizon and its higher-spin generalizations.
In this method, it is important that matter fields 
in the black hole background behave as if they
are a collection of  two-dimensional conformal
fields near the horizon in $(t,r_*)$ coordinates,
where $r_* $ is the tortoise coordinate. 
The conformal invariance near the horizon 
has been emphasized 
 \cite{conformal} in  understanding 
the black hole entropy by the near horizon 
Virasoro symmetries.  
We also show that the response of physical quantities
to a conformal transformation gives the correct value
of the Hawking radiation.

There are several derivations of Hawking radiation
and all of them take into account 
the quantum effect in the 
black hole backgrounds  in various
ways.  In the original derivation \cite{Hawking1}, the radiation
appears as a result of the particle interpretation 
of the quantum wave propagation in the black hole
background.
An understanding of the Hawking radiation in the
field theory language was first given by 
Christensen and Fulling \cite{CF}.
In this calculation,  the energy 
flux of the radiation is given by using the
conservation law of the energy-momentum tensor
and the quantum trace anomaly.

Recently Robinson and Wilczek \cite{RW}  showed that 
the energy flux of the Hawking radiation 
can be fixed by the value of the gravitational
anomaly at the horizon. 
Here 
the universality of Hawking radiation 
is connected to the universality of the
gravitational anomaly.  
This work was further generalized by us to obtain the 
flux of charge from charged black holes 
by considering the gauge
anomaly in addition to the gravitational one \cite{IUW1}.
More recently the full thermal spectrum of the 
Hawking radiation 
was given by  T. Morita, H. Umetsu and the present author
by calculating the anomalies for the
higher-spin currents \cite{IMU4}.

In this review, we summarize these recent
developments to understand  the universal
behavior of the Hawking radiation. 
In the next section, we first emphasize the 
emergence of the conformal
symmetry for matter fields near the horizon.
In section 3, we then explain how to calculate the 
expectation value of the 
energy-momentum tensor  in two different
methods.  The first method is to calculate the 
response of the holomorphic energy-momentum 
tensor to a conformal transformation, 
and the second one is a method using 
the gravitational  anomaly. In section 4, 
we generalize these two methods to 
higher-spin currents and obtain the full thermal
spectrum of the Hawking radiation.
Finally we conclude in section 5.
\section{Near Horizon Conformal Symmetry}
For simplicity, we consider a charged scalar field in a 
$d=4$ Reisnner-Nordstrom 
black hole background with a metric 
$ds^2= f(r)dt^2 -f(r)^{-1}dr^2 -r^2 d\Omega^2$
and $f(r)=1-2M/r+Q^2/r^2.$
The scalar field is  minimally coupled and the action is given by 
\begin{equation}
S = \int d^4 x \sqrt{-g} \{
g^{\mu \nu} (\partial_\mu + ieA_\mu) \phi^*
(\partial_\nu - ie A_\nu) \phi -m^2 \phi^*\phi -V(\phi)
\}.
\end{equation}
By decomposing the scalar field into partial waves as 
$\phi=\sum_{l,m} \phi_{lm}(r,t)Y_{lm}(\Omega)$, 
 the action becomes an infinite set of two-dimensional
 fields in the $(t, r_*)$ coordinate as 
\begin{eqnarray}
 S &=& \sum_{l,m} \int dt dr_* r^2(r_*)
  \big[
  |(\partial_t-ieA_t)\phi_{lm}|^2 - |\partial_{r_*}\phi_{lm}|^2 
 \nonumber \\
    && -f(r(r_*)) \left( 
  m^2 |\phi_{lm}|^2 + \frac{l(l+1)}{r^2}|\phi_{lm}|^2 + V(\Phi_{lm})
 \right)
  \big].
  \label{decomposedaction}
\end{eqnarray}
Here $r_*$ is the tortoise coordinate.
When we consider a matter field with finite energy 
at $r=\infty$, the $d=2$ kinetic term behaves regularly near the 
horizon while the potential term becomes negligible because
of the damping coefficient $f(r)$ near the horizon.
Hence the $d=4$ scalar field can be considered as an 
infinite set of $d=2$ conformal fields near the horizon
in the $(t, r_*)$ coordinate. 
This is the most essential property
to the universality of the Hawking
radiation.

The emergence of the two-dimensionality and
the conformal symmetries near the horizon
is very general and the following derivations
of the Hawking radiation are generally applicable
to higher dimensional cases. 
When the black hole is rotating, the system can be 
reduced to a charged black hole by 
the dimensional reduction and the analysis becomes
identical to the case of the charged black holes \cite{IUW2}.
It is also true for any higher dimensional black holes or 
other black objects and the gravitational anomaly
method has been applied to various cases \cite{others}.

Two comments are in order.
First, in order to evaluate the Hawking fluxes at infinity, we 
need to extrapolate the fluxes near the horizon to
infinity, and the thermal black-body 
radiation is distorted 
by the grey body factor by the effect of 
scattering of waves in the potential.
There are two sources for the potential.
The first is the original potential term in the action
(\ref{decomposedaction}).
The second  is 
the $r$-dependence of the volume factor. 
Its main effect is to relate the integrated flux 
over the sphere in $d=4$ with the $d=2$ outgoing flux, but
in addition to it, this gives an effective potential 
proportional to $1/r^3.$
In this review, we neglect the effects of the potential
(grey body factor)
and mainly consider the universal behavior of radiation
determined by the information at the horizon.

Next I would like to comment on another possible violation of
the universality of Hawking radiation.
The damping coefficient $f(r)$ of the potential 
terms in (\ref{decomposedaction}) comes from the
relation  $\partial_r = f(r)^{-1} \partial_{r_*}$.
Hence, if the action has  higher-dimensional 
operators containing higher derivatives
with respect to $r$, they become more dominant
than the ordinary kinetic terms.
For example, a term $M_{UV}^{-2}|\partial_{r}^2 \phi|^2$ is 
enhanced near the horizon by a factor  
$f(r)^{-2}$ and  becomes more relevant
than the ordinary kinetic terms, 
if the enhancing factor dominates
the low-energy suppression factor $(T_{BH}/M_{UV})^2$.
Here $T_{BH}$ is the typical energy scale 
of Hawking radiation at infinity.
This indicates that the local field
theory based on a low-energy approximation
is no longer valid near the horizon.
This is an interesting issue, 
but we leave it for a future problem.
\section{Universality of Hawking Energy Flux}
In this section, we discuss why the energy flux can be 
universally determined only by the information at the horizon.
We first obtain the Hawking energy flux by calculating the response
of the holomorphic energy-momentum tensor to a conformal
transformation from the Kruskal to the Schwarzschild coordinates.
In the second subsection, we review the derivation of the Hawking
energy flux based on the gravitational anomaly at the horizon.
\subsection{Response to conformal transformations}
For simplicity we consider a neutral  field in 
the Schwarzschild black hole with $f(r)=1-2M/r$. The tortoise
coordinate describes the region outside the horizon and 
is defined by  $dr_*=dr/f(r)$. It behaves as
$r_* \sim r$ at infinity and takes $-\infty$ near the horizon.

There are two important coordinate systems; the  Schwarzschild
light cone coordinates labeled by the outgoing $u=t-r_*$ and 
ingoing $v=t+r_*$ coordinates, and the Kruskal coordinates
defined by $U=-\exp(-\kappa u)$ and $V=\exp(\kappa v)$
where $\kappa=1/4M$ is the surface gravity.
The Schwarzschild coordinates are convenient
to describe the physics
at infinity, while the Kruskal coordinates are
more appropriate to discuss the physics observed 
by an infalling observer near the horizon. 
Hawking radiation emanates from 
the future horizon $U \sim 0$ of the black hole,
and the regularity condition imposed on the 
physical quantities observed by an infalling
observer determines the boundary condition for the
 Unruh vacuum (with an additional condition for the ingoing 
modes)\cite{Unruh}.

Consider a reduced two-dimensional system
describing  one partial wave.
In the approximations with the grey body factor neglected, 
the two-dimensional energy-momentum tensor for each partial wave 
is conserved
$\nabla_\mu T^\mu_\nu=0$  and its trace receives
the Weyl anomaly as $T^\mu_\mu= c \hbar R/24 \pi$ where
$c=1$ for a boson and $c=1/2$ for a fermion.
In the curved space-time with 
$ds^2=e^{\varphi(u,v)}dudv$, these two equations 
enable us to define the  holomorphic
energy-momentum tensor as
\begin{equation}
t_{uu}(u)=  T_{uu}(u,v) -\frac{c}{24 \pi}
\left( \partial_u^2 \varphi - \frac{1}{2}(\partial_u \varphi)^2
\right),
\label{holomorphicEM}
\end{equation}
which does not depend on the $v$ coordinate.
This holomorphic energy-momentum tensor is no longer an ordinary
tensor 
and under a coordinate transformation from $u$ to $U(u)$, 
it transforms as
\begin{equation}
t_{UU}(U)= \left( \frac{1}{\kappa U} \right)^2
\left( t_{uu}(u) + \frac{c}{24 \pi} 
\{ 
U,u
\}
\right), \ \ \ \{U,u \}= \frac{U^{'''}}{U'} -\frac{3}{2}
\left( \frac{U''}{U'} \right)^2.
\label{SD1}
\end{equation}
It is important to note that $t_{uu}$ and $T_{uu}$ coincide
at $r=\infty$ because the conformal
factor vanishes there.  Hence the value of $t_{uu}(u)$ at 
$r=\infty$ gives the outgoing flux from the black hole.
The value can be fixed by imposing the condition
that the physical quantities should be regular
for an infalling observer  in the Kruskal coordinate. 
In order that $t_{UU}$ should behave regularly near the 
future horizon at $U \sim 0$, $t_{uu}$ must 
cancel the term $c \{U,u\}/24\pi$. Hence we get the value of 
the outgoing energy flux by the value of the Schwarzian
derivative as
\begin{equation}
T_{uu} = - \frac{c\hbar}{24\pi} \{U,u\}= \frac{c \hbar \kappa^2}{48 \pi}.
\end{equation}
If we further assume that the ingoing flux vanishes
at infinity $T_{vv}=0$ for the Unruh vacuum, 
we get the asymptotic energy flux at $r=\infty$ as 
$T^r_t= T_{uu}-T_{vv}=c\hbar \kappa^2/48\pi.$
This is equal to the integral of the thermal distribution
\begin{equation}
T^r_t= \int_0^\infty \frac{d\omega}{2\pi}\frac{\hbar \omega}{e^{\beta \omega}\mp 1}
= \frac{c\hbar \kappa^2}{48\pi}
\label{energyflux}
\end{equation}
where  the inverse
temperature $\beta$ is given by $\beta=2\pi/ \kappa.$

The method  can generalized to obtain an energy flux in 
a charged black hole \cite{IMU2}. 
In this case, the conservation of energy-momentum tensor
is modified to $\nabla_\mu T^\mu_\nu =F_{\mu \nu}J^\mu$ in the presence
of the electric field and the $U(1)$ current $J^\mu$, and 
 the holomorphic
energy-momentum tensor is changed accordingly.
The energy flux can be  obtained by considering
the response of the holomorphic energy-momentum tensor 
to the coordinate transformation
and an additional gauge transformation.
\subsection{Gravitational anomaly method}
In this subsection we review the recent derivation
of the Hawking radiation based on the gravitational
anomaly at the horizon \cite{RW,IUW1}. The basic idea is the following.
As we saw, each partial wave of 
a matter field behaves as a free field in 
the two-dimensional $(t,r_*)$ coordinates, and 
the outgoing (ingoing) modes  correspond to the
right (left) moving modes in $d=2.$
Classically nothing can escape from the black hole
and the ingoing modes at the horizon cannot affect
the physics outside of the black hole.
So they can be neglected classically.
But if we neglect them the theory
becomes chiral, and quantum mechanically
it would break the gauge and general coordinate invariance
through the quantum anomalies.
Of course, the underlying theory is not anomalous
and the classically irrelevant ingoing modes 
become relevant quantum mechanically.
This gives the flux of the Hawking radiation.

In order to understand the Hawking radiation
as stated above, it is 
  suitable to consider the path integral
formulation and investigate the Ward-Takahashi identities
in the black hole background as in ref. \refcite{RW,IUW1}.
In this formulation, the WT identities are written
in terms of the consistent currents while the boundary
conditions are imposed on the covariant currents, and
we need to relate these two different types of currents.
Here we instead take an equivalent but simpler method using only the 
covariant currents given by Banerjee and Kulkarni \cite{Banerjee}.

First consider the $U(1)$ current in a charged black hole.
If we neglect the ingoing modes at horizon, the $U(1)$
current becomes anomalous. We further simplify the situation
and assume that the ingoing modes are neglected
everywhere. This can be justified if there is no
scattering (no grey body factor) and by imposing
the Unruh type boundary condition that the ingoing
flux should vanish at infinity. Then we have the anomalous equation
for the covariant form of the $U(1)$ current 
\begin{equation}
\nabla_\mu J^\mu = \frac{\hbar}{4\pi} \epsilon_{\mu \nu} F^{\mu \nu}.
\label{gaugeanomaly}
\end{equation}
In the static  background, it can be easily soloved as
\begin{equation}
\langle J^r(r) \rangle = c_H + \frac{\hbar}{2\pi}\left( A_t(r) -A_t(r) \right).
\end{equation}
Here the integration constant $c_H$ 
gives the value of the covariant outgoing current 
at the horizon.
The regularity condition at the future horizon imposes
that $c_H=0$, otherwise the outgoing current 
$J_U=-J_u/\kappa U$ observed by an
 infalling observer diverges at $U=0.$
Hence the $U(1)$ flux at infinity can be obtained as 
$c_O=J^r(r=\infty)=-\hbar A_t(r_+)/2\pi =\hbar Q/2\pi r_+$ where $r_+$
is the position of the outer horizon. 

The flux is determined  by the gauge potential at the horizon.
This universal behavior is obtained in the anomaly method
because the r.h.s. of  (\ref{gaugeanomaly})
is written as a total derivative and, by integrating
(\ref{gaugeanomaly}), the current can be obtained by the 
boundary condition at the horizon only.

It can be generalized to the energy flux as well.
We  consider the covariant energy-momentum tensor
with the covariant form of the gravitational anomaly;
\begin{equation}
\nabla^\mu T_{\mu\nu}=F_{\mu\nu}J^\mu + \frac{\hbar}{96\pi}
\epsilon_{\mu\nu}\partial^\mu R.
\label{gravitationalanomaly}
\end{equation}
It is important to note again that the r.h.s. of this equation
is written as a total derivative and the energy-momentum
tensor can be solved as
\begin{equation}
T^r_t=a_H+\int_{r_+}^r \partial_r 
\left[
c_O A_t +\frac{\hbar}{4\pi}A_t^2 + \frac{\hbar}{96\pi}(ff''-\frac{(f')^2}{2})
\right].
\end{equation}
The regularity condition imposes $a_H=0$, and
the asymptotic outgoing energy flux
can be determined by the value of the anomaly at the horizon as
\begin{equation}
a_O=T^r_t(r=\infty )= \frac{\hbar Q^2}{4 \pi r_+^2} +\frac{\hbar \pi}{12\beta^2}.
\end{equation}

\section{Higher-spin Fluxes and Anomalies}
In the previous section, we have calculated the energy flux
based on two different methods. But the energy flux is a partial information 
of the thermal radiation which can be obtained as an
integral (\ref{energyflux}).
It would be desirable if 
we can obtain the full thermal spectrum of the Hawking radiation,
or equivalently all the moments 
\begin{equation}
\int_0^\infty \frac{d\omega}{2\pi}
\frac{\hbar \omega^{2n-1}}{e^{\beta \omega} \mp 1}.  
\label{moments}
\end{equation}
In the following we show that they can be fully reproduced
by the anomaly method or the conformal transformation method.

\subsection{Conformal transformation properties of  higher-spin currents}
The energy-flux of the Hawking radiation was reproduced 
by calculating the Schwarzian derivative of the holomorphic
energy-momentum tensor  in (\ref{SD1}). 
This method was generalized by us \cite{IMU2} to
obtain all the moments of  (\ref{moments}).

Here I will explain the case of a neutral scalar field, but
 generalizations to  charged fields or fermion fields
are straightforward. 
First consider a 4-th rank current 
constructed from a scalar field as
 $J^{(1,3)}= - :\partial_u \phi \partial_u^3 \phi (u):$.
This is regularized by the point splitting regularization, and
a standard calculation shows that
under a conformal transformation $u \rightarrow U(u)=-e^{-\kappa u}$
it transforms as
\begin{eqnarray}
:\partial_u \phi \partial_u^3 \phi (u): &=&
\kappa^4 U^2 :\partial_U\phi^{(U)}  \partial_U \phi^{(U)}:
 + 3 \kappa^4 U^3 :\partial_U\phi^{(U)}  \partial_U^2 \phi^{(U)}:
 \nonumber \\
&&+ \kappa^4 U^4 :\partial_U\phi^{(U)}  \partial_U^3 \phi^{(U)}:
-\frac{\hbar}{480\pi} \{ U, u\}_{(1,3)}
\label{spin3holo}
\end{eqnarray}
where the generalized Schwarzian derivative is given by
\begin{equation}
\{ U, u\}_{(1,3)} =\frac{6U'''}{U'}-20 \left( \frac{U'''}{U'} \right)^2
-45 \left( \frac{U''}{U'} \right)^4 +90 \frac{(U'')^2U'''}{(U')^3} 
-30 \frac{U'''U''}{(U')^2}.
\end{equation}
 The regularity condition at the future horizon imposes again that
 the first three quantities in the r.h.s. of (\ref{spin3holo})
 should vanish at $U=0$,
 and the spin-4 flux at infinity is given by 
 $\langle J^{(1,3)}(r=\infty) \rangle
  =\hbar \{ U, u\}_{(1,3)}/480\pi = \hbar \kappa^4/480\pi $,
 which agrees with the $n=2$ moment of the bosonic case (\ref{moments}).
Here note that another spin-3 current $:\partial_u^2 \phi \partial_u^2 \phi (u):$
gives the same answer for the asymptotic flux and hence
any linear combination of them does not change the final result, if 
appropriately normalized. 

It can be straightforwardly generalized to any higher-spin currents,
but instead of calculating them one by one, it is easier to 
calculate  Schwarzian derivatives for
 a generating function of the higher-spin currents
\begin{equation}
J(u,u+a)= :\partial_u \phi (u) \partial_u \phi (u+a): = \sum_{n=0}^\infty 
\frac{a^n}{n!} :\partial_u \phi (u) \partial_u^{n+1} \phi (u): .
\end{equation}
Under a transformation $u\rightarrow U(u)$, it transforms as
\begin{equation}
J^{(U)}(U(u),U(u+a)) 
= e^{\kappa a} \left( \frac{1}{\kappa U}\right)^2  
\left( J(u,u+a)  - \hbar A_b(U,u) \right). 
\label{fluxgenerating}
\end{equation}
where $A_b(U,u)$ is given by
\begin{equation}
A_b(U,u) = - \frac{\partial_u U(u) \partial_u U(u+a)}{4\pi(U(u)-U(u+a))^2}
+ \frac{1}{4\pi a^2}= -\frac{\kappa^2}{16\pi \sinh^2(\kappa a/2)}
+ \frac{1}{4\pi a^2}.
\end{equation}
The regularity condition at $U=0$ determines the value
of the flux as 
$\langle J(u,u+a) \rangle= \hbar A_b(U,u)$.
By expanding it with respect to $a$,
we get the right answer for the flux of general higher-spin currents;
$\langle : (1)^{n-1 }\partial_u \phi \partial_u^{2n-1} \phi (u):
\rangle = B_n \kappa^{2n}/8\pi n$, where $B_n$ is the Bernoulli number.
$A_b(U,u)$ is nothing but the temperature dependent part
of the thermal Green function of a scalar field.

\subsection{Higher-Spin Gauge Anomaly}
The gravitational anomaly method can be also 
generalized to the higher-spins and reproduce the
correct thermal fluxes\cite{IMU4}. 
For this purpose, we need to 
obtain  generalized equations of the gravitational
anomaly (\ref{gravitationalanomaly}) to higher-spin currents.
The next simplest example is the conservation equation for
the spin-3 current.  For the spin-3 current constructed 
from a fermionic field, it is given  by

\begin{align}
 \nabla_\mu {J^{(3)\mu}}_{\nu\rho}
 =& \left( - F_{\nu\mu} {J^{(2)\mu}}_\rho
 -\frac{1}{16}\nabla_\nu (R  J^{(1)}_\rho ) 
 + (\nu \leftrightarrow \rho)
 \right)  
%
%
%
 + \frac{1}{16}g_{\nu\rho}\nabla_\mu \left(R J^{(1)\mu} \right) \nonumber \\
 &+ \frac{\hbar}{96\pi}
 \left( 
 \epsilon_{\nu \sigma} \nabla^\sigma \nabla_\mu {F^\mu}_{\rho} 
 + \epsilon_{\rho \sigma} \nabla^\sigma \nabla_\mu {F^\mu}_{\nu}
 - g_{\nu\rho} \epsilon_{\alpha \sigma} \nabla^\sigma 
 \nabla_\mu F^{\mu\alpha}
 \right).
 \label{higherspinanomaly}
\end{align} 
The first line of r.h.s. is the classical violation 
@in the presence of electric and gravitational background, 
while the second line is the quantum anomaly. 
In the present context  the transformation generated by
the spin-3 current is not gauged and it is not the anomaly
associated with the gauge symmetry. 
But since it has the same quantum origin as the gravitational anomaly
for the energy-momentum tensor,  we call it a higher-spin
gauge anomaly here. An interesting property of the r.h.s. is that
both of the classical and the quantum violations are
written as  total derivatives as is the case for the energy-momentum
tensor. Hence 
the asymptotic flux from a black hole,
which can be obtained by integrating the equation
and imposing the regularity condition at the horizon,
is determined only by the information at the horizon as
\begin{equation}
c_o^{(3)}=-\frac{\hbar \kappa^2 A_t(r_+)}{24\pi} -\frac{\hbar A_t(r_+)^3}{6\pi}
= \frac{\hbar \kappa^2 Q}{24\pi r_+} + \frac{\hbar}{6\pi} \left( \frac{Q}{r_+} \right)^3.
\end{equation}
Similar  calculations can be done for higher-spin currents.
It is interesting that the higher-spin anomalies can be also written 
as  total derivatives.  

Finally we  summarize  in four steps how to obtain the  equation like 
(\ref{higherspinanomaly}): 
\\
(i) Regularize higher-spin currents covariantly on the light-cone  with $v$ fixed.\\
(ii)  Regularize the higher-spin currents holomorphically. \\
(iii) Compare these two and obtain a relation like (\ref{holomorphicEM}) for higher-spin currents. \\
(iv) Covariantize the relation. \\
With these four steps, we can obtain classical and quantum  violations of 
the conservation equations for the higher-spin currents
in the presence of the electric and gravitational backgrounds.
For further details, please refer to  ref. \refcite{IMU4}.
Furthermore we can obtain the quantum violation of the 
classically traceless symmetric currents. For example, the spin-3 current
has a trace anomaly
\footnote{The authors of ref.\refcite{Bonora} 
state that the trace anomaly for spin 4 current  
is cohomologically trivial.  
It is interesting to see
whether the higher-spin anomalies 
can be  absorbed by  redefining the currents.  } 
\begin{equation}
J^{(3)\mu}_{\mu\nu} = \frac{\hbar(c_L+c_R)}{24\pi} \nabla_\mu F^\mu_\nu
\end{equation}
where $c_L$ and $c_R$ are left (right) central 
charges.
\section{Conclusions and Discussions}
Hawking radiation is determined only by the information at the 
horizon, if  grey body factor is neglected. The universality
can be explained in various ways, but it can be
beautifully explained in the anomaly method. 
Namely the quantum anomaly is written 
as a total derivative and the regularity condition 
determines the value of the asymptotic flux
by the value of the surface term of the anomaly
at the horizon.
This is reminiscent of the topological effects of instantons in the
gauge theories. In the present case, the horizon
plays a role of a boundary and the regularity condition
there relates the asymptotic flux to the information at the horizon.

Another interesting issue is a possible violation of the universality,
in the context of the information paradox, 
due to the enhancement of higher derivative terms very close
to the horizon, as discussed in section 2. We would like to
come back to this problem in  future.

\section*{Acknowledgments}
I would like to thank the organizers of the workshop for 
a wonderful atmosphere of the meeting.
This talk is based on collaborations with
T. Morita, H. Umetsu  and F. Wilczek,
to whom I am grateful for stimulating discussions and collaborations.

\end{document}